\documentclass[aps,prb,twocolumn,psfig]{revtex4}
\usepackage{epsfig}
\newcommand{\be}{\begin{equation}}
\newcommand{\ee}{\end{equation}}
\newcommand{\bea}{\begin{eqnarray}}
\newcommand{\eea}{\end{eqnarray}}

\newcommand{\modification}{}
\newcommand{\gergo}{}

\begin{document}
\title{Existence of two-channel Kondo regime for tunneling impurities with 
resonant scattering}
\author{G. Zar\'and}
\address{
Department of Theoretical Physics, 
Budapest University of Technology and Economics, Budafoki \'ut 8., H-1521 Hungary
}
\date{\today}

\begin{abstract}
Dynamical tunneling systems have been proposed earlier to display a two-channel Kondo effect,
the orbital index of the particle playing the role of a pseudospin in the equivalent 
Kondo problem, and the spin being a silent channel index. However, as shown 
by Aleiner {\em et al.} [Phys. Rev. Lett. {\bf 86}, 2629 (2001)], the predicted two-channel 
Kondo behavior can never be observed in the weak coupling regime, where the tunneling 
induced splitting of the levels of the tunneling system always dominates the physics.
Here we show that the above scenario changes completely in the strong coupling regime, 
where - as a non-perturbative analysis reveals -  the two-channel Kondo regime can easily 
be reached. We show that a tunneling systems end up quite  naturally in this regime if the 
conduction electrons are scattered by {\em resonant scattering} off the tunneling 
impurity, {\gergo and we also speculate about the possible origins of such a  resonant scattering.}
\end{abstract} 
\pacs{72.10.Fk,72.15.Qm}
\maketitle


\section{Introduction}
There are  a number of somewhat mysterious low-temperature transport anomalies in disordered 
point contacts,\cite{RalphReview98,Upadhyay97,Keijsers96,Balashkin98,Balashkin01} 
structurally disordered single crystals and disordered alloys,\cite{cichorek04,cichorek01,katayama87,brandt82}, 
which have not been 
explained satisfactorily. In some cases the observed anomalies display power law behavior
\cite{RalphReview98,cichorek_unpub} ($\sim T^{1/2}$ or $V^{1/2}$), sometimes they show universal scaling 
properties,\cite{RalphReview98} but they have the common feature that  all of them seem  to be 
related to the presence of  some dynamical impurities.   Although very attractive, electron-electron 
interactions fail to explain the absence of this zero bias anomaly in point contacts with 
strong static disorder,\cite{AltshulerComment} and  the complete absence of magnetic 
field dependence in some  experiments.\cite{Upadhyay97,cichorek01} 

\begin{figure}
\begin{center}
\epsfxsize7.0cm
\epsfbox{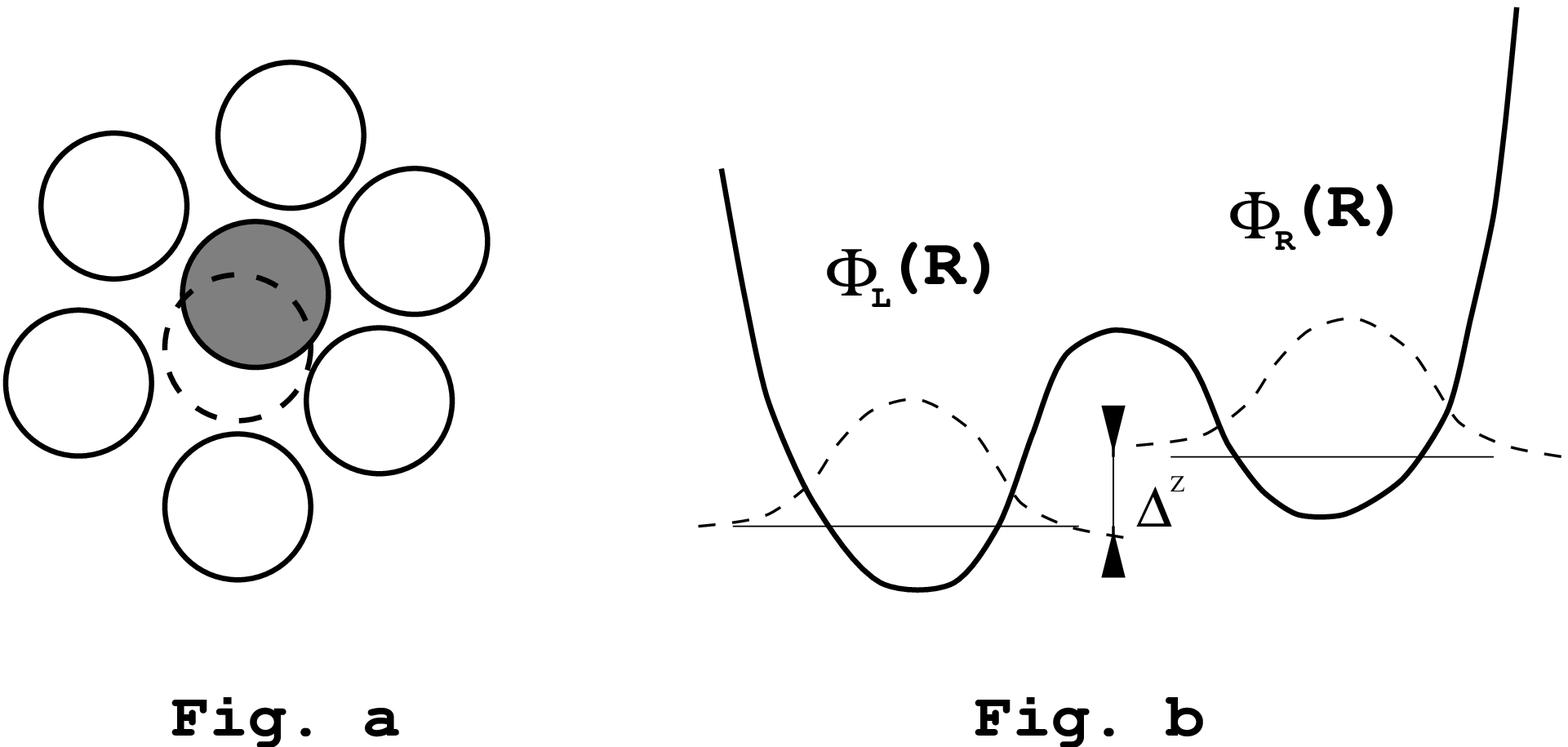}
\end{center}
\caption{\label{fig:dwp}
Sketch of the tunneling  system. The tunneling system is probably formed by a single atom 
in an amorphous region (Fig.~a). The tunneling atom moves in an effective double well potential (Fig.~b). 
}
\end{figure}

Much before the above-mentioned experimental results, Vlad\'ar and Zawadowski proposed that dynamical two-level systems
- abundant in amorphous regions\cite{Black} -
could lead to an orbital  Kondo effect.\cite{vladzaw,Kondo} They considered the motion of an ion in an effective 
double well 
potential, depicted in Fig.~\ref{fig:dwp}, interacting with free conduction electrons. They assumed a simple 
potential scattering interaction between the atom and the conduction electrons,  and  
derived the following effective Hamiltonian for low temperatures, where 
the atom moves by tunneling between the two minima of the potential well:
\begin{eqnarray}
H &= &- {\Delta_0\over 2} \tau_x - {\Delta_z\over 2} \tau_z  
\nonumber \\
& + & {v_z} \sum_\sigma  \tau_z (\psi^\dagger_{e\sigma} \psi_{o\sigma}+\psi^\dagger_{o\sigma} \psi_{e\sigma})
\nonumber \\
&+ &{v_x} \sum_\sigma  \tau_x (\psi^\dagger_{e\sigma} \psi_{e\sigma}-\psi^\dagger_{o\sigma} \psi_{o\sigma})\;.
\label{eq:H_VZ}
\end{eqnarray}
Here the Pauli matrices $\tau_i$ describe the motion of the particle in the double well potential, 
with $\tau_z=\pm1$ corresponding to the left and right potential wells, $\Delta_0$ the tunneling, 
and $\Delta_z$ the asymmetry of the potential (see Fig.~\ref{fig:dwp}). The operators  
$\psi^\dagger_{e/o\;\sigma}$ create  conduction electrons in some  even/odd angular momentum channel 
($s$ and $p$, {\em e.g.}),\cite{vladzaw} 
respectively,  and are defined as 
$$
\psi^\dagger_{e/o\;\sigma}=\int_{-D_0}^{D_0} d\xi\;\psi^\dagger(\xi)_{e/o;\sigma}\;.
$$
Here $D_0$ is a high energy cutoff discussed later and 
the $\psi^\dagger(\xi)_{e/o\;\sigma}$'s satisfy canonical anti-commutation relations:
$$
\{\psi^\dagger(\xi)_{\mu\sigma}, \psi(\xi')_{\mu'\sigma'}\} = \delta(\xi-\xi')\delta_{\mu\mu'}
\delta_{\sigma\sigma'}\;.
$$
This corresponds to a normalization of the fields $ \psi_{e/o,\sigma}$ in Eq.~(\ref{eq:H_VZ}) such that
the imaginary time propagators at $T=0$ temperature behave asymptotically as 
\be
\langle T_\tau \psi_{\mu,\sigma}(\tau) \psi^\dagger_{\mu',\sigma'}(0)\rangle = {\delta_{\mu\mu'}\over \tau}\;.
\label{eq:normalize}
\ee
The dimensionless couplings $v_x$ and $v_z$ in Eq.~(\ref{eq:H_VZ}) and the splitting 
$\Delta_0$ have also been  estimated by Vlad\'ar and Zawadowski. Assuming a simple 
$s$-wave scattering $U$ off the tunneling impurity they found 
\bea
\Delta_0 & \approx & \hbar \omega_0 e^{-\lambda}\;,
\label{eq:tunamp}
\\
v_x & \approx & U\varrho_0  {(k_F d)^2\over 24}  {\Delta_0\over V_B}\;,\\
v_z & \approx & U\varrho_0  {k_F d \over \sqrt{3}} \;,
\eea
where $\varrho_0$ denotes the density of states of the conduction electrons at the Fermi energy, 
$d$ is the tunneling distance, $k_F$ the Fermi momentum, $V_B$ is the height of the tunnel barrier, 
and $\lambda$ the Gamow factor. The attempt frequency $\omega_0$ is typically somewhat less than the 
Debye temperature, and is typically in the range $\omega_0\sim 100 \;{\rm K}$ assuming a tunneling atom of 
mass $M\sim 50\; m_p$, with $m_p$ the proton mass.
 Vlad\'ar and Zawadowski also obtained the perturbative scaling equations 
for the model defined by Eq.~(\ref{eq:H_VZ}) and showed that the couplings $v_i$ and 
the dimensionless tunneling  $\tilde \Delta_0 \equiv \Delta_0/D$ satisfy the following 
differential equations
\bea
{d v_x\over dl} = 4 v_y v_z - 8 \; v_x (v_y^2 + v_z^2)\;,
\label{eq:pert_v}
\\
{d \tilde \Delta_0 \over dl} = (1 - 8 (v^2_z+v_y^2)) {\tilde \Delta}_0\;,
\label{eq:pert_Delta}
\eea
where $l=\ln(D_0/D)$ denotes the scaling variable with $D$ the 
running cutoff (energy scale).  
The other equations are obtained by cyclic permutation from Eqs.~(\ref{eq:pert_v}) and (\ref{eq:pert_Delta}). 
The coupling $v_y$ is absent in the original Hamiltonian, it is, however, generated by the scaling 
procedure. 
Clearly, 
Eq.~(\ref{eq:pert_v}) - apart from a prefactor in the second term - is 
that of the anisotropic Kondo model,\cite{vladzaw}  and generates  a Kondo effect 
at the energy scale $\sim T_K$, the Kondo temperature, 
where the running coupling constants become of the order of unity.
This Kondo effect is, however, strikingly different from the ordinary Kondo effect, where the
low temperature physics is that of a Fermi liquid below $T_K$.  
In the present case spin is conserved in the course of scattering, and 
the  two spin channels  give rise  to a singular two-channel Kondo 
behavior   below $T_K$,  {\gergo provided that the  splitting of the  two levels 
can be neglected}.\cite{Cox} 
The anomalous properties of this two-channel Kondo 
state (resistivity $\sim T^{1/2}$, $V/T$ scaling {\em etc.}) have been proposed to  explain the experimentally 
observed anomalies.\cite{RalphReview98}  However, the tunneling amplitude 
$\Delta_0$ is always a relevant variable, and it ultimately kills orbital fluctuations 
(and thus the Kondo effect) of the tunneling system below a characteristic energy scale 
$\Delta_0^*$. If this energy scale $\Delta_0^*$
is larger than $T_K$ then the anomalous properties of the 
two-channel Kondo fixed point cannot be observed.
Vlad\'ar and Zawadowski  argued that - for a symmetrical tunneling system - there is a parameter range where  
$\Delta_0^*\ll T_K$, implying that there is a temperature window  where 
the physics is { dominated by the Kondo  effect}.

Unfortunately, Vlad\'ar and Zawadowski assumed in their analysis that the scaling equations above 
are valid at all energy scales below the Fermi energy $E_F$. However, as pointed out by Aleiner 
{\em et al.},\cite{aleiner} this assumption is wrong: Electrons with energy $\xi \gg \omega_0$ 
follow the tunneling particle {\em adiabatically},\cite{Kagan}  and do not give contribution 
to the vertex renormalization. As a  consequence, equations (\ref{eq:pert_v}) and (\ref{eq:pert_Delta}) 
become trivial for $\omega_0 < D < E_F$: 
\be
\left.\begin{array}{c}
{d v_i/ dl} = 0\phantom{n} \\
{d \tilde \Delta_i /  dl} = 1 \phantom{n} 
\end{array}
\right\}\mbox{   for}\phantom{n} \omega_0 < D < E_F\;.
\label{eq:trivial_scaling}
\ee
Therefore, as pointed out by Aleiner {\em et al.}, Eqs.~(\ref{eq:pert_v}) and (\ref{eq:pert_Delta}) 
must be solved with the initial conditions $D= \omega_0$,  $\tilde \Delta_0 \sim  e^{-\lambda} > 
v_x \sim (\varrho_0 U) (k_F d)^2 e^{-\lambda}$. Since in the perturbative 
regime $\tilde \Delta_0$ grows always faster that $v_x$, 
$\Delta_0^*$ is always larger than $T_K$, and the two-channel Kondo effect can never be observed 
for $v_z \ll1$, in contrast to the original conclusions of Vlad\'ar and Zawadowski.

{\gergo In this paper we first show that the arguments of Aleiner {\em et al.}  hold only in the weak coupling 
regime, $v_z \ll1$. In the strong coupling regime, $v_z \sim 1$ one must treat the coupling $v_z$ 
non-perturbatively and use the scaling equations originally derived by  
 Vlad\'ar,  Zim\'anyi,
and Zawadowski in Ref.~\onlinecite{VladZimZaw}, that treat $v_z$ {\em exactly} while    
handling the small coupling $v_x$ and the tunneling only in leading order. The analysis based on these equations 
clearly shows that there exist a critical value,  $v_z=v_{z,c}\equiv 1/\pi$. Above this  
 so-called Emery-Kivelson line, \cite{EmeryKivelson}  $v_x$ {\em typically  wins} over $\Delta_0$, {\em i.e.,}, 
the renormalized splitting is smaller then 
the Kondo temperature,   $T_K > \Delta_0^*$,  and thus there is usually  a wide temperature range 
where the physics of the  tunneling system is {\em dominated by the two-channel Kondo fixed point}.  

Then we show that a  possible and probably the most  natural way to get into this regime, 
 is to have a tunneling 
atom that also acts as a {\em resonant scatterer} at the Fermi 
energy.\footnote{{\gergo {\em I.e.}, the atom itself that moves has an electron  structure with resonant scattering 
at the Fermi energy, even if its position is fixed.}}
Indeed, as already observed  in the original work of Vlad\'ar and Zawadowski, it is known from experiments 
that the value of $v_z$ can be {\em large}, suggestive of {\em resonant scattering} from the tunneling 
systems, whatever they are. As we shall see, in this case it is not enough to start from the 
simple potential scattering model studied in Refs.~\onlinecite{vladzaw,aleiner}  and one must take into 
account the dynamical motion of the internal levels of the tunneling atom. As we show later, 
due to this resonant scattering, one easily gets outside the range of validity of the 
perturbative equations considered by Aleiner {\em et al.}, Eqs.~(\ref{eq:pert_v}) 
and (\ref{eq:pert_Delta}), and ends up in the strong coupling regime.}

\begin{figure}
\begin{center}
\includegraphics[width=7cm]{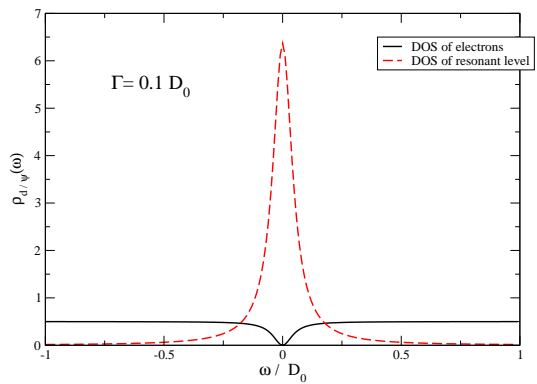}
\end{center}
\caption{\label{fig:dos}
Density of states (spectral function)  of a resonant level $d$  and the conduction electron operator 
$\psi$ hybridizing with it. The conduction electrons' spectral function is suppressed 
While $\varrho_d$ displays a huge resonance at the Fermi energy.
}
\end{figure}

To show why  resonant scattering  is of primary  importance, let us consider a simple toy model 
describing an atom with a single resonant level at the Fermi energy,
\be
H = \sum_{\sigma, {\bf k}}\xi({\bf k}) \; c^\dagger_{{\bf k},\sigma}  c_{{\bf k},\sigma}
+ V\sum_{\sigma, {\bf k}}  (  c^\dagger_{{\bf k},\sigma}  d_\sigma + {\rm h.c.})\;.
\ee
Here $d_\sigma$ is the annihilation operator describing the level,  $V$ is the hybridization, 
and $ c^\dagger_{{\bf k},\sigma} $ creates a conduction electron with spin $\sigma$, momentum ${\bf k}$, and 
energy $\xi({\bf k})$. This model can be trivially solved, and we can compute  the spectral functions 
$\varrho_d(\omega)$ and  $\varrho_\psi(\omega)$ of the $d$-level and the 
fermionic field at the impurity site, $\psi_\sigma \equiv \sum_{\bf k} c_{{\bf k},\sigma}$,
\bea
\varrho_d (\omega)  = {1\over 2\pi} {\Gamma\over \omega^2 + \Gamma^2/4}\;,\\
\varrho_\psi (\omega)  = \varrho_0  {\omega^2\over \omega^2 + \Gamma^2/4}\;.
\eea
Here $\varrho_0$ is the density of states of the conduction electrons at the Fermi energy, 
and $\Gamma = 2 \pi \varrho_0 V^2$ is the width of the resonance.   Clearly, if we now try to move the 
atom, it will then couple to {\em its own resonant level} $d$ the most strongly, 
which is centered at the atom, and not to the rest of the conduction electrons,
 which are sitting at neighboring ions in reality. 
Even  for a not too  narrow resonance with $\Gamma\sim 1\; {\rm eV}$,
the density of states of the resonant level at the Fermi energy is large, $\sim 1/\Gamma$, 
 while that of the conduction electrons hybridizing with the atom actually vanishes
at the Fermi energy (see Fig.~\ref{fig:dos}).
This large increase in the density of states is what immediately  
pushes the system in the strong coupling regime,
$v_z> v_{z,c}$, where the two-channel Kondo behavior prevails, 
{\gergo provided that the  resonance is narrow enough. For typical parameters we obtain that,  
the resonance should be narrower than about $0.1 \;E_F - 0.01 \; E_F \sim  1000- 10000 \;{\rm K}$.}

{\gergo We  emphasize though that  the above resonance does not have to be narrower than $\omega_0$: 
As we discussed earlier, only electrons  with energy $|\omega|<\omega_0$  are unable  to follow the 
motion of the atom at a time scale 
$\sim 1/\omega_0$. These latter are exactly the electrons that are responsible for the Kondo effect. However, 
these electrons  only need a time scale $\sim 1/\Gamma$ to notice the increased scattering strength 
at the Fermi energy, {\em i.e.}, all these 'slow' electrons see an increased scattering 
strength from the atom. This simple picture can be readily verified\cite{Zarunpub}  through a path integral 
treatment similar to that of Ref.~\onlinecite{aleiner2} as well as by the extension of the 
the diagrammatic  approach of Refs.~\onlinecite{aleiner,borda}.}
{\gergo For a resonance  narrower than $\omega_0$ our analysis must be slightly modified, and 
the vertex renormalization only occurs in the range $D<\Gamma$, since electrons with 
energy $\omega > \Gamma$ do not couple  strongly to the atom.}

{\gergo As we show later, the increase in the coupling constant can simply be understood
as a matrix element effect: 
the wave functions of the conduction electrons 
at the Fermi energy (where the resonance is located) have an {\em increased amplitude} 
at the tunneling atom's position, and therefore the motion of the atom couples more strongly 
to the conduction electrons.}

\begin{figure}
\begin{center}
\epsfxsize7.0cm
\epsfbox{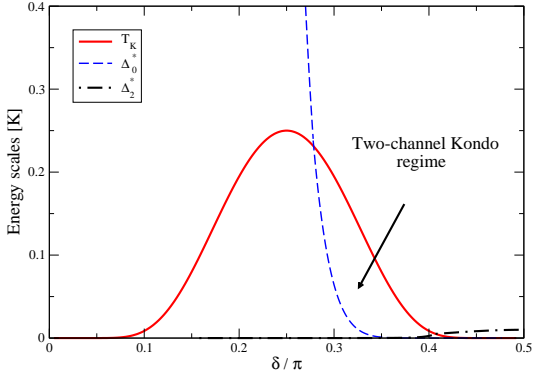}
\end{center}
\caption{\label{fig:phasediag}
Phase diagram of the tunneling system as a function of the phase shift $\delta = {\rm artan}(v_z \pi)$. 
The arrow indicates the two-channel Kondo regime. The symbol  $\Delta_0^*$ denotes  the energy scale 
where the tunneling becomes dominant, while  $\Delta_2^*$ is the energy scale where a special  
two-electron scattering process becomes important. We have used $\omega_0 \approx 100{\rm K}$, 
$D_0\sim 10^5 \;{\rm K}$, 
$\Delta_0 = 10 {\rm K}$, and $v_x\approx 0.1$ as bare parameters.}
\end{figure}

It is thus the structure of a resonant 
level itself which is ultimately responsible for the orbital Kondo effect in the above scenario, 
and this resonance cannot be replaced by a simple potential scatterer in a free electron gas.\cite{footnote} 
The technical reason for this is that in a simple potential scattering model
 there is always  a term $\sim\sum_{\sigma,\mu=e,o} v_{0,\mu} \psi^\dagger_{\mu\sigma} \psi_{\mu\sigma}$, which 
is large, cannot be neglected, leads to a suppression of the conduction electron's density of states, 
and ultimately reduces the effective value of $v_z$.  
As we shall see, the effective Hamiltonian for this resonant level finally takes also the form, 
Eq.~(\ref{eq:H_VZ}), however, the couplings will be related to some atomic orbitals, and no large
term $\sim   v_0  \sum_{\sigma,\mu}  \psi^\dagger_{\mu\sigma} \psi_{\mu\sigma}$ appears.
Therefore we can use the non-perturbative scaling equations valid for all values of 
$v_z$, derived longtime ago by Zim\'anyi, Vlad\'ar, and Zawadowski in Ref.~\onlinecite{VladZimZaw} 
 to construct the phase diagram of our model. The  summary of this analysis is shown in Fig.~\ref{fig:phasediag}.
The bell-shaped line in Fig.~\ref{fig:phasediag} shows the estimated value of the Kondo 
temperature, $T_K$, while $\Delta_0^*$ is the renormalized tunneling amplitude. 
In Fig.~\ref{fig:phasediag} we also show another energy scale, $\Delta_2^*$, associated with 
a two-electron scattering process.\cite{MoustakasFisher} Below this energy scale the two-channel
 Kondo behavior is also suppressed. As one can see from Fig.~\ref{fig:phasediag}, there is an extended 
regime dominated by the  two-channel Kondo fixed point. The size of this region increases if we decrease the value 
of $v_x$ and $\Delta_0$, however, then $T_K$ is also shifted towards smaller temperatures.
As we discussed  above, the tunneling system ends up immediately in this regime if there is 
{\gergo a sufficiently narrow} resonant scattering on the tunneling atom.

{\gergo We would also like to mention that another possibility to get to the strong coupling 
regime is to {\em increase the tunneling distance} and have a broader resonance at the Fermi 
energy.  This might be possible for very light tunneling impurities, possibly by hydrogen:
hydrogen has typically resonant scattering in the $s$-channel and it can possibly 
tunnel over a large distance in a metal. Zero bias anomalies have indeed been observed 
in hydrogen doped palladium point contacts  where the anomaly is clearly related to the presence of 
hydrogen.\cite{Mihaly}}

The rest of the paper is structured as follows: In 
Section~\ref{sec:micro} we shall construct a microscopic model that enables us to treat the 
resonance properly and map this problem to the original Vlad\'ar-Zawadowski model.
We then analyze the scaling equations and construct the phase diagram of this model in section 
\ref{sec:scaling}. Finally, in the concluding section we shall speculate about possible 
candidates producing a resonant scattering at the Fermi energy.

\section{Microscopic model}
\label{sec:micro}

As a first step in constructing a microscopic model for a tunneling system coupled to the electrons,
 we write  the Hamiltonian in a first quantized form as 
\bea
H & = & - {1\over 2M} \Delta_R  
- \sum_i {1\over 2m} \Delta_{i} 
\label{H_real}
\\
& + & U({\bf R}) 
+ \sum_i V({\bf r}_i-{\bf R})\;
+ \sum_{j\ne0} \sum_{i} V({\bf r}_i- {\bf R}_j)\;, 
\nonumber
\eea
where ${\bf R}$ denotes the coordinate of the tunneling atom, and  ${\bf r}_i$'s
are the coordinates of the conduction electrons. 
We treat the motion of the tunneling atom quantum mechanically, however, we assume 
that all other atoms are immobile, and their position ${\bf R}_j$ ($j=1,\dots$) is a constant
classical variable. The tunneling atom has a mass $M$ and moves in the double well
 potential $U({\bf R}) $ formed by the rest of the ions. 
In the Hamiltonian (\ref{H_real}) we also assumed that electrons form a
 Fermi liquid and thus neglected  electron-electron interaction, which is supposed 
to be included in the effective electron-ion interaction  potential $V({\bf r}- {\bf R})$ 
at the Hartree or Hartree-Fock level. 

To simplify this Hamiltonian, as a first step, we rewrite 
Eq.~(\ref{H_real}) as $H=H_0+H_{\rm int}$ with 
\bea
H_0 & = & - {1\over 2M} \Delta_R  + U({\bf R}) 
\label{H_real2}
\\
& - & \sum_i {1\over 2m} \Delta_{i} 
+ \sum_{j=0,\dots}  \sum_{i} V({\bf r}_i-{\bf R}_j)\;, 
\nonumber
\\
H_{\rm int} & = &  \sum_i \left( V({\bf r}_i - {\bf R})- V({\bf r}_i-{\bf R}_0)\right)\;, 
\label{eq:H_int}
\eea
where ${\bf R}_0$ is a somewhat arbitrary 'reference position' of the ion, which is chosen 
to minimize  the interaction part $H_{\rm int}$. A natural choice is, of course to choose 
${\bf R}_0$ to correspond to the maximum of the barrier in $U({\bf R})$. Clearly, the 
non-interacting part $H_0$ can be diagonalized. 

To make further progress, we shall adopt a tight-binding scheme for the atom. Though this approach 
 is  justified by the observation that the tunneling impurity moves in a small {\em cavity}, our analysis 
does not  rely on it, and our conclusions are independent of this approximation. 
To arrive at a tight-binding Hamiltonian,  we first solve the atomic Schr\"odinger equation:
\be 
\left( -  {1\over 2m} \Delta_{\bf r} +    V({\bf r}-{\bf R}_0)\right) 
\varphi_\mu ({\bf r}) = \epsilon_\mu \;
\varphi_\mu ({\bf r}) \;,
\ee
where $\epsilon_\mu$ labels the atomic levels. The states $\varphi_\mu $ above are atomic eigenstates
centered at the tunneling atom. As a next step, we solve the Schr\"odinger equation for the 
rest of the conduction electrons  without making any approximation, 
\be
\left( -  {1\over 2m} \Delta_{\bf r} +   
\sum_{j\ne0} V({\bf r}-{\bf R}_j)\right) \eta_n ({\bf r}) = \xi_n \;
\eta_n ({\bf r}) \;.
\ee
Then, using the wave functions $\varphi_\mu$ and $\eta_n$, we can 
compute the appropriate overlap matrix elements and construct the following tight 
binding Hamiltonian for the conduction electrons:
\bea
H_0^{\rm el} &=& \sum_{\mu,\sigma} \epsilon_\mu d^\dagger_{\mu,\sigma} d_{\mu,\sigma}
+ \sum_{n,\sigma} \xi_n c^\dagger_{n,\sigma} c_{n,\sigma}
\nonumber \\
&+ & \sum_{\mu,n,\sigma} (t_{n,\mu} c^\dagger_{n,\sigma} d_{\mu,\sigma} + h.c.)\;,
\eea
where $c^\dagger_{n,\sigma}$ denotes the creation operator of a conduction electron 
in the (extended) state $n$ with spin $\sigma$, 
$d^\dagger_{\mu,\sigma}$ creates a conduction electron at the atomic orbital $\mu$, and 
$t_{n,\mu}$ denotes the corresponding hopping matrix element. Since the interaction part 
$H_{\rm int}$ is large only at the position of the tunneling impurity, 
we can integrate out all electrons  $c^\dagger_{n,\sigma}$ and  arrive at the following imaginary time 
effective action for the $d$-levels:
\be
S^0_{\rm eff} = - \sum_{\mu,\mu',\sigma} \int_0^\beta d\tau  \int_0^\beta d\tau'  
{\bar d}_{\mu\sigma}(\tau){\cal G}_{\mu\mu'}^{-1}(\tau-\tau') { d}_{\mu'\sigma}(\tau')\;,
\ee
where ${\cal G}_{\mu\mu'}$ denotes the local propagator of the $d$-levels,
which can be expressed  in Fourier space as
\be
{\cal G}_{\mu\mu'}^{-1}(i\omega_n) = i\omega_n - \epsilon_\mu \delta_{\mu\mu'}
 - \sum_n t^*_{n,\mu}  {1\over i\omega_n - \xi_n} t_{n,\mu'}\;.
\ee 
This Green's function can be written  even more conveniently  in a spectral representation 
as 
\be
{\cal G}_{\mu\mu'}(i\omega_n) = \int_{-\infty}^\infty d\omega \; 
 {\varrho_{\mu\mu'}(\omega)\over i\omega_n - \omega}\;.
\ee
It is this spectral function $\varrho_{\mu\mu'}(\omega)$ 
which contains the resonance discussed in the previous section, and which 
solely  determines the low temperature properties of the tunneling system.

In the following we shall consider the simplest case, where the tunneling system is fully 
symmetrical.  We first diagonalize the Hamiltonian of the tunneling atom, 
\be
\left(- {1\over 2M} \Delta_R  +  U({\bf R}) \right) \Phi_\alpha({\bf R}) 
= E_\alpha \Phi_\alpha({\bf R})\;. 
\label{eq:eigen_Phi}
\ee
In principle, we could formulate our theory by 
keeping all  levels of the tunneling particle,\cite{ZarZawPRL}
However, at low enough temperatures only the two lowest lying even and odd states, 
$\Phi_{e}$ and $\Phi_o$ matter, and the role of all other eigenstates is to reduce 
the electronic cutoff to a value of the order of the Debye frequency, $D_0\to \omega_0$.\cite{aleiner,borda}
Therefore, in the following we shall keep only these two states. To obtain the 
usual tunneling form that occurs in  Eq.~(\ref{eq:H_VZ}), we introduce the left and right states
$\Phi_{\tau=\pm}$
\be
\Phi_\pm \equiv {1\over \sqrt{2}}(\Phi_e-\Phi_o)\;,
\ee
which transform into each-other under reflection, and rewrite
the Hamiltonian Eq.~(\ref{eq:eigen_Phi}) in this restricted  subspace  as 
\be 
H_0^{\rm tun} = -{\Delta_0\over 2} \tau_x\;,
\ee
with $\tau_x$ the Pauli matrix.  The tunneling amplitude $\Delta_0 = E_o - E_e$
is approximately given by  Eq.~(\ref{eq:tunamp}).\cite{vladzaw}

Having diagonalized  the non-interacting part (\ref{H_real2}) of the Hamiltonian, we now 
turn to  the analysis of the interaction part. First, we shall simplify our treatment by 
keeping only those two even and odd  electronic states   $d_\mu$ ($\mu=e,o$), which couple the 
most strongly to the tunneling particle.
In a real system, these correspond to the most strongly coupled $d$ and 
$p$-states, or  $s$ and  $p$ states. In case of a $p$-state, {\em e.g.}, 
it is clear that the $p$-state aligned along the tunneling axis couples the most strongly 
to the motion of the atom. Due to the assumed spatial symmetry of the Hamiltonian, 
the corresponding spectral function is diagonal in the index $\mu$, $\varrho_{\mu\mu'} = 
\delta_{\mu\mu'} \varrho_\mu$. We shall assume furthermore that at least one of the 
spectral densities, say   $\varrho_o$ contains a {\em resonance} of width $\Gamma$, 
while the other one is approximately constant, $\varrho_e\approx \varrho_0\sim 1/E_F$
\bea
\varrho_o (\omega) & = & {1\over 2\pi} {\Gamma\over \omega^2 + \Gamma^2/4}\;,
\\
\varrho_e (\omega) & = & \varrho_0\;.
\eea
To construct the interaction part of the Hamiltonian in this restricted basis, 
we have to compute matrix elements of $H_{\rm int}$ given by Eq.~(\ref{eq:H_int}). 
Assuming that the tunneling distance is small, we can approximate $H_{\rm int}$ as
\be 
H_{\rm int} \approx -\sum_{q=x,y,z} R_q {\partial V\over \partial r_q}
+ \frac 12 \sum_{p,q=x,y,z} R_p R_q {\partial^2 V\over \partial r_p \partial r_q}
+ \dots\;,
\ee
and then compute the appropriate matrix elements to obtain the effective Hamiltonian 
as
\bea
&& H_{\rm int}   \approx   A (|\Phi_e\rangle \langle \Phi_o| + |\Phi_o\rangle \langle \Phi_e|)
 (d^\dagger_{e\sigma} d_{o\sigma} + d^\dagger_{o\sigma} d_{e\sigma}) 
\nonumber 
\\
&+& (B_e |\Phi_e\rangle \langle \Phi_e| + B_o |\Phi_o\rangle \langle \Phi_o| )
 (C_e d^\dagger_{e\sigma} d_{e\sigma} + C_o d^\dagger_{o\sigma} d_{o\sigma})\;, 
\nonumber 
\eea 
where summation is assumed over repeated spin indices, and the constants $A$, $B_{e/o}$ and 
$C_{e/o}$ are given by the following integrals:
\bea
A = -  \langle\Phi_e| R_z| \Phi_o\rangle 
          \langle \varphi_e| {\partial V({\bf r})\over \partial r_z}| \varphi_o \rangle \;,
\\
B_{e/o} = {1\over 2} \langle \Phi_{e/o}| R_z^2 |\Phi_{e/o}\rangle\;,
\\
C_{e/o} =  \langle \varphi_{e/o}|  {\partial^2 V({\bf r})\over ({\partial r_z})^2} |\varphi_{e/o}\rangle\;.
\eea 
This Hamiltonian can easily brought to the form, Eq.~(\ref{eq:H_VZ}), if we notice that 
only electronic excitations with energies $<\omega_0$  contribute to the 
orbital Kondo correlations.\cite{aleiner} Therefore, if the width of the resonance is broader than the 
attempt frequency, $\Gamma>  \omega_0\sim 100K \sim 0.01 {\rm eV}$, then we can introduce the 
new fermionic fields  normalized  according to Eq.~(\ref{eq:normalize})
\bea 
d_{e\sigma} \rightarrow \psi_{e\sigma} \equiv d_{e\sigma}/\sqrt{\varrho_{e}}\;,
d_{o\sigma} \rightarrow \psi_{o\sigma} \equiv d_{o\sigma}/\sqrt{\varrho_{o}}\;,
\eea
where $\varrho_{e}$ and $\varrho_{o}$ denote the density of states at the Fermi energy 
in the even and odd channels, respectively.  In terms of these fields, the interaction 
Hamiltonian takes on the form  Eq.~(\ref{eq:H_VZ}), with the  couplings $v_x$ and $v_z$ given by: 
\bea
v_z &=&  A \sqrt{\varrho_e \varrho_o} \;,
\nonumber
\\
v_x &=& {1\over 4} (B_e - B_o) (C_e \varrho_e - C_o \varrho_o)\;.
\label{eq:couplings_res}
\eea
There is two more terms that appear in addition to these two terms, both of which are 
small: One is a simple potential  scattering term, $\sim (\psi^\dagger_e \psi_e - \psi^\dagger_o \psi_o )$ 
that only renormalizes the  spectral densities $\varrho_o$ and $\varrho_e$, and can be eliminated 
by a simple counterterm  procedure. The other term is proportional to 
$\sim \tau_x (\psi^\dagger_e \psi_e + \psi^\dagger_o \psi_o )$,   and gives a small 
renormalization of the double well potential $U({\bf R})$.\cite{borda} This term can be taken into account at the 
Hartree level and is of no importance. In the following we shall therefore keep only 
the  two couplings $v_x$ and $v_z$  in Eq.~(\ref{eq:couplings_res}).

{\modification Note that the above procedure of integrating out the electrons $c_{n,\sigma}$ and then 
rescaling the couplings is just a technical trick to extract the dimensionless couplings and to derive 
quickly the scaling equations. However, one can proceed in the usual way as
and obtain  the dimensionless couplings by analysis the structure of the dimensionless vertex functions, 
as in Refs.~[\onlinecite{borda,Nozieres}].}  

The couplings $v_z$ and $v_x$ can be estimated along similar lines as in Refs.~[\onlinecite{vladzaw,Cox}] 
and one obtains:
\bea
v_z & \approx & \sqrt{\varrho_e \varrho_o} 
\langle \varphi_e| d{\partial V({\bf r})\over \partial r_z}| \varphi_o \rangle\;,
\\
v_x & \approx & - \Delta_0 {\lambda\over 16} {1\over V_B}  \varrho_o 
 \langle \varphi_{o}| d^2 {\partial^2 V({\bf r})\over ({\partial r_z})^2} |\varphi_{o}\rangle\;,
\label{v_x_estim}
\eea
where in the second equation we
 assumed a quartic double well potential with barrier height $V_B$, 
and displayed only the contribution of the odd channel. The constant $d$ in the previous equations 
denotes the tunneling distance.
A similar but smaller contribution is given by the non-resonant
even channel. 
 Remarkably, the matrix  elements above are expressed in terms of  the atomic (tight binding) orbitals 
of the tunneling atom. They can be  easily estimated assuming a simple Coulomb interaction, $V({\bf r})= -e^2/r$,
and hydrogen-like wave functions. Taking the $1s$ and $2p$ orbitals,
{\em e.g.},  we find 
\bea 
\langle 2p\; | {\partial V({\bf r})\over \partial r_z}|\;1s \rangle = {e^2\over a_0} {4\over 27 \sqrt{2}\; a_0}
\approx 5.4 \;{{\rm eV}\over \AA}\;,
\\
\langle 2p\;| {\partial^2 V({\bf r})\over ({\partial r_z})^2} |\;2p\rangle = - {e^2\over a_0} {1\over 40\; a_0^2}
\approx -2.43 \;{{\rm eV}\over \AA^2}\;,
%
\eea
where $a_0$ is the Bohr radius.

It is not difficult  to show that even for a wide  resonance with  $\Gamma\sim 1000-10000 \;{\rm K}$
and $d\sim 0.3 \AA$,  $v_x$ is about the same  as the dimensionless tunneling 
$\Delta_0 /\omega_0$. More importantly, however, 
$v_z$ is proportional to $1/\sqrt{\Gamma}$. Therefore, if the conduction electrons scatter resonantly off
the tunneling atom,  then the value of $v_z$ can become large, and eventually become larger 
than the critical value corresponding to the Emery-Kivelson line mentioned in the introduction, 
$v_{z,c} = 1/\pi$.  For typical parameters this transition takes place where the width of the resonance is around 
$\Gamma \sim 1000-10000 \; {\rm K}$. 
  
{\modification Throughout this paper, we shall assume that $\Gamma > \omega_0$, {\em i.e.}, that the resonance is 
relatively broad compared to typical frequencies related to atomic motion.
Under these conditions, the cutoff energy is given by $\omega_0$ and the couplings between the 
tunneling system and the electrons can be safely  approximated by their values at the Fermi energy.} 

\begin{figure}
\begin{center}
\includegraphics[width=7cm,clip]{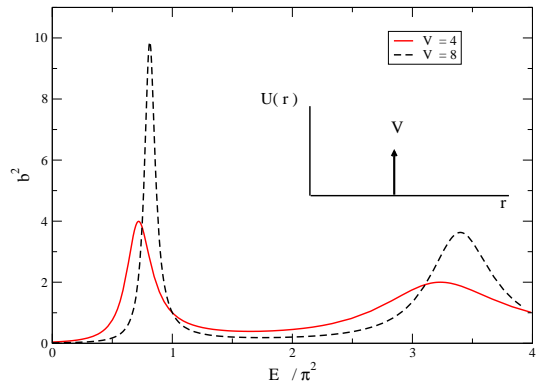}
\end{center}
\caption{\label{fig:amplitude}
Amplitude $b^2$ of the 
scattering state of the conduction electons at the origin in a simple scattering state model.
The amplitude of the scattering state at the origin increases as the resonance gets sharper and sharper.  
The inset shows the potential producing the resonance.}
\end{figure}

{\modification The above large increase in the effective couplings can be understood as a simple  matrix 
element effect, and is related to the well-known structure of scattering states: To clarify this point, let us 
consider the textbook example of scattering states in the $s$-channel of a simple spherically symmetric 
potential scatterer,  $U(r) = V\;\delta(r-r_0)$. The scattering wave functions in this simple case 
take the form $\sin(k r + \delta)/r$ and  $b\;\sin(k r)/r$ for $r>r_0$ and 
$r<r_0$, respectively, with $k$ the radial momentum of the electrons and $\delta$ the scattering phase shift.
The amplitude $b^2$ of the wave function inside the sphere is given by the expression,
\be
b^2 = \frac{k^2}{k^2 + V^2 \sin^2(k) + 2 k V \sin(2k)}\;,
\ee
Where we used units of $r_0 = 2 m = \hbar = 1$. For $V\equiv 0 $ the amplitude of the wave function is simply 
one. However, for larger values of $V$ a resonance appears, and $b^2$ displays a sharp peak as a function 
of  energy (Fig.~\ref{fig:amplitude}). 
It is a trivial matter to show that the amplitude $b$ of the resonance  is simply related to the width $\Gamma$
of the resonance, $b\sim 1/\sqrt{\Gamma}$, and becomes larger and larger for sharper and sharper resonances.
This large factor $b$ shows up in the local density of states and also any matrix element computed in terms 
of the appropriately normalized scattering states and results in an increase of all couplings, 
provided that the resonance appears at the Fermi energy.}

\section{Non-perturbative scaling analysis}
\label{sec:scaling}

We have seen in the previous section that for resonant scattering from the  tunneling impurity the 
coupling $v_z$ can be very large. In this case the perturbative scaling equations, Eqs.~(\ref{eq:pert_v}-\ref{eq:pert_Delta})
 are insufficient  and a new approach is needed.
Fortunately, for ${\tilde \Delta}_0,\;v_x\ll 1$ one can construct scaling equations which are
{\em non-perturbative} in the coupling $v_z$ by generalizing the computations of Yuval and 
Anderson.\cite{VladZimZaw,YuvalAnderson} 
In this limit one obtains the following scaling equations\cite{VladZimZaw} 
\bea
{d{\tilde \Delta}_0 \over dl}= \left[1-8\left({\delta\over\pi}\right)^2\right]\; {\tilde\Delta}_0\;,
\label{eq:Delta_0_VZZ}
\\
{d v_x  \over dl} = 4{\delta\over \pi} v_y - 8\left({\delta\over\pi}\right)^2 v_x\;,
\label{eq:v_x_VZZ}
\\
{d v_y  \over dl} = 4{\delta\over \pi} v_x - 8\left({\delta\over\pi}\right)^2 v_y\;,
\label{eq:v_z_VZZ}
\eea
with the phase shift $\delta$  defined as
$\delta = {\rm arctan}( \pi v_z)\;$. Here we neglected terms of order ${\cal O}(v_x^2, v_y^2, \tilde \Delta_0^2)$. These terms 
give rise to a renormalization of $\delta$ in the strong coupling regime, and slightly change the numerical values of 
the various energy scales we estimate, but do not affect their overall scale and the  picture obtained using 
the above equations.
Equations (\ref{eq:v_x_VZZ}) and (\ref{eq:v_z_VZZ}) can be rewritten in terms of the average coupling,
$v_\perp \equiv (v_x + v_y)/2$, and the asymmetry, $v_- \equiv (v_x - v_y)/2$ as
\bea
{d v_\perp  \over dl} & = & \left[ 4{\delta\over \pi}  - 8\left({\delta\over\pi}\right)^2 \right] v_\perp\;,
\label{v_perp_VZZ}
\\
{d v_-  \over dl} & = & - \left [4{\delta\over \pi}  + 8\left({\delta\over\pi}\right)^2 \right] 
v_-\;.
\label{v_-_VZZ}
\eea
In other words, the anisotropy  $v_x\ne v_y$ is {\em irrelevant}, while  the average 
coupling $v_\perp$ is relevant. Therefore the two couplings $v_x$ and $v_y$ become 
rapidly equal in the initial stage of the scaling.
Eqs.~(\ref{v_perp_VZZ}) and (\ref{v_-_VZZ})   are only valid below the scale 
$D \approx \omega_0$; above this energy scale the couplings  $v_x$, $v_y$ remain unrenormalized, while 
 ${\tilde \Delta}_0$ transforms according to its engineering dimension and satisfies  Eq.~(\ref{eq:trivial_scaling}).
Therefore Eqs.~(\ref{v_perp_VZZ}) and (\ref{v_-_VZZ})  must  be solved with the initial conditions
$D = \omega_0$, ${\tilde \Delta}_0(\omega_0)= \Delta_0/\omega_0 \sim e^{-\lambda}$, 
$v_y(\omega_0) =0$ and $v_x(\omega_0)$ given by  Eq.~(\ref{v_x_estim}).
 
From Eqs.~(\ref{eq:Delta_0_VZZ}) and (\ref{eq:v_x_VZZ}) 
we see that the value $v_z=1/\pi$ corresponding to $\delta = \pi/4$ is very special:
 For $\delta < \pi/4$ the scaling 
dimension of ${\tilde \Delta}_0$ is always larger than that of $v_x$, and therefore  
tunneling is always more relevant than assisted tunneling. As a result, the motion of 
the tunneling system freezes out at an energy scale $\Delta_0^* \ll \Delta_0$
\be
\Delta_0^* = \Delta^*_{0,>}  \equiv  \Delta_0 \left({\Delta_0\over \omega_0}\right)^\alpha\;,\phantom{nn}
\alpha = {8\left({\delta\over\pi}\right)^2 \over 1-8\left({\delta\over\pi}\right)^2}\;,
\label{eq:Delta^*_0,>}
\ee
where the assisted tunneling is still small compared to unity, and thus the system
is far from the two-channel Kondo fixed point.  

Eq.~(\ref{eq:Delta^*_0,>}) only makes sense if $\Delta_0^*> T_K$. This is always the case for
$\delta<\pi/4$.
If, however, $\delta> \pi/4$ then the physics can become very different. In this case $v_x$ grows 
{\em faster} than ${\tilde \Delta}_0$, so $v_x\sim 1$ can be satisfied first. The condition 
$v_x \sim1$ defines the so-called Kondo temperature:
\be
T_K \approx \omega_0 \left({v_x\over 2}\right)^\gamma\;,\phantom{nn}
\gamma^{-1} = {   4{\delta\over \pi}  - 8\left({\delta\over\pi}\right)^2}\;,
\ee
where the renormalized  tunneling satisfies
\be
{ \Delta_0}(T_K) \approx  T_K {2{ \Delta}_0\over \omega_0 \;v_x} \left({v_x\over 2}\right)^\beta\;,
\phantom{nn}
\beta = {{4{\delta\over \pi} -1} \over 4{\delta\over \pi}\left(1 - 2{\delta\over \pi}\right)}\;.
\label{Delta_0(T_K)}
\ee
From this equation it is obvious that for $\delta>\pi/4$ the effective tunneling amplitude 
is usually still small at the energy scale $T\sim T_K$ compared to the temperature itself.
Below the Kondo  energy $T_K$  the scaling equations are governed by the two-channel Kondo 
fixed point. There the tunneling is still relevant and has a scaling dimension $1/2$ and therefore its 
scaling equation must be replaced by\cite{Cox}
\be
{d{\tilde \Delta}_0\over dl}  = {1\over 2} \;{\tilde \Delta}_0\;, \phantom{nnn}(D< T_K)\;.
\ee
Integrating this  equation we obtain for the renormalized tunneling amplitude
for the case $\Delta_0^*<T_K$,
\bea
\Delta_{0,<}^* & \approx & T_K \left({2{\tilde \Delta}_0\over v_x}\right)^2\;
\left({ v_x\over 2}\right)^{2\beta} \ll T_K \;.
\eea

From our discussions immediately  follows that in the tunneling regime for a symmetrical tunneling center 
with $v_z>v_{z,c}$ there is typically {\em  a  large temperature window}, 
$\Delta_0^* <  T < T_K$,  
 where the two-channel Kondo fixed point 
rules and the non-Fermi liquid properties such as the $\sim \sqrt{T}$ resistivity 
anomaly associated with the two-channel Kondo fixed point should  be manifest.
We have to emphasize,  however, that below  the scale $\Delta_0^*$ the physics becomes again 
that of a boring Fermi liquid.  The corresponding crossover lines 
were sketched in Fig.~\ref{fig:phasediag} for typical parameters of the tunneling system.
As one can see in the figure, there is a large region in the parameter space, which is governed 
by the two-channel Kondo behavior. This region increases even further for smaller values of $\tilde \Delta_0 \sim v_x$, 
however, it also shifts towards smaller values of $T_K$.

Before we conclude this section, let us discuss  another important issue, raised by Moustakas 
and Fisher.\cite{MoustakasFisher} Moustakas and Fisher observed that at the two-channel Kondo fixed 
point  a special two-electron scattering  of the form [$\psi_\pm = (\psi_e \pm \psi_o)/\sqrt{2}$],
\be 
H_2 = {\tilde \Delta_2\over D_0} \tau_x 
(\psi^\dagger_{+\uparrow} \psi^\dagger_{+\downarrow} \psi_{-\downarrow}\psi_{-\uparrow} + h.c.)
\ee
 is also relevant, and that for $\delta > \pi/4$ 
this operator is {\em more relevant} than the tunneling $\tilde \Delta_0$  discussed above.
The coupling $\tilde \Delta_2$ above denotes a dimensionless coupling
constant, and its bare value can be estimated to be around   $v_x$, while the energy scale 
$D_0$ is of the order of the Fermi energy $E_F$. 
Below the energy scale $D=\omega_0$ this operator satisfies the scaling equation:
\be
{d{\tilde \Delta}_2 \over dl}= \Bigl(-1+ 8 {\delta\over\pi} -8\left({\delta\over\pi}\right)^2\Bigr)\; 
{\tilde\Delta}_2\;.
\ee
Fortunately, this operator is {\em irrelevant} at high energies, $\omega_0<D$, where its scaling dimension 
is simply $-1$,
\be
{d{\tilde \Delta}_2 \over dl} = - {\tilde\Delta}_2\;, \phantom{nnn}( D>\omega_0)\;.
\ee
 As a result, by the time we get into the regime $D<\omega_0$ this two-electron process
 is reduced by  a factor $\omega_0/D_0 \sim 10^{-3}$ compared to $v_x$.

Below $\omega_0$ this process becomes relevant, and generates a cross-over to a Fermi liquid 
at an  energy scale $\Delta_2^*$. Similar to  $\Delta_0^*$, we have to distinguish two possibilities. 
For larger values of $\delta$,   $T_K<\Delta_2^*$, and $\Delta_2^*$ is given by the following 
expression: 
\bea
\Delta_2^* & \approx & \Delta_{2,>}^* \equiv \omega_0 \left({\omega_0\over E_F} \tilde \Delta_2\right)^\kappa\;,  
\\
\kappa^{-1} & = &  8 {\delta\over\pi} -8\left({\delta\over\pi}\right)^2 -1\;.
\eea
However, this formula is not correct if  $T_K>\Delta_2^*$, since 
$\tilde \Delta_2$ has scaling dimension $1/2$  below the Kondo temperature.\cite{MoustakasFisher}
 Carrying out an analysis similar to the case
of $ \Delta_0^*<T_K$ we find in this regime that
\bea 
\Delta_2^* & \approx & \Delta_{2,<}^* \equiv 
\omega_0 \left({\omega_0\over E_F} \tilde \Delta_2\right)^2 \left({v_x\over 2}\right)^\rho \;,
\nonumber
\\
\rho & = & {3 + 16 \left({\delta\over\pi}\right)^2 - 16 {\delta\over\pi}\over 
4 {\delta\over\pi} - 8 \left({\delta\over\pi}\right)^2}\;.
\eea
The corresponding crossover line is also shown in Fig.~\ref{fig:phasediag}. As one can see, 
$\Delta_2$ does not play a significant role in the regime where $T_K$ is the largest, however, 
for larger values of $\delta$ it is indeed $\Delta_2$ that provides an infra-red cutoff 
for the two-channel Kondo behavior.

\section{Conclusion}
\label{sec:conclusion}

In the present paper we constructed a theory which describes the  low temperature behavior of  
a tunneling particle with resonant scattering. We have shown that the resonance plays a 
crucial role in the physics of the tunneling center:
While the local density of states of the resonant atomic  
state has a huge peak inversely proportional to the width of the resonance, 
$\varrho_d\sim 1/\Gamma$,  the density of states of the corresponding conduction electrons 
is strongly suppressed, and thus the tunneling atom couples the most efficiently to 
its own electronic state. Despite  this complication, after a proper treatment of the electron-tunneling particle interaction,
 we arrive at  the original  model of Vlad\'ar and Zawadowski. However, our treatment does not rely on the 
free electron approximation, the various coupling constants can be large and they are 
determined by atomic integrals. 

As we have shown, the resonant scattering can completely change the physics of the tunneling impurity, 
and push it in the non-perturbative regime. While in the weak coupling regime the tunneling 
always intervenes before we enter the two-channel Kondo regime, in this 
non-perturbative regime the two-channel Kondo behavior usually dominates over the
spontaneous tunneling over a wide temperature range of possibly several decades for physically 
relevant parameters. We emphasize again that a model where the tunneling center interacts with 
some local interaction with a free electron gas is unable to capture this behavior.\cite{aleiner,footnote} 
The physical reason  for this is simply that electrons are {\em not free}. Even if we do not move the 
tunneling atom, the electron's wave function is adjusted to the atomic potential of the tunneling impurity. 
This effect has been heuristically taken into account in Ref.~\onlinecite{borda}, where only the {\em change}
in the scattering potential has been considered as a perturbation, without justifying this approach. 
Also, the simple potential scattering model is unable to  take into account the dynamics 
(retardation effects) of the electronic states at the tunneling impurity itself, which one actually 
tries to eliminate from the theory.

It is not quite clear what the origin of such a resonant level could be in practice. 
In the point contact measurements, one natural candidate would be hydrogen. Hydrogen  
is small enough, can diffuse into the substrate, and  is known to have a phase shift $\pi/2$ 
in the $s$-channel.\cite{Hydrogen} [This phase shift, which just characterizes the atomic scattering off
a Hydrogen ion is not to be confused with the phase shift $\delta$ corresponding to $v_z$.]
Moreover, hydrogen is light: This implies that the cutoff $\omega_0$ 
can be in the range of $\omega_0 \sim 1000\;{\rm K}$. As a result, all temperature scales
can be about an order of magnitude larger for hydrogen  than the ones in Fig.~\ref{fig:phasediag}, 
and $T_K$ can be easily 
in the range of a few  Kelvins. Being light,  hydrogen can tunnel 
over relatively large distances, $d\sim 1\;\AA$, implying that $v_z$ is presumably large 
even if the resonance is very broad. In fact, in the point contact experiments it is very difficult 
to exclude the presence of  hydrogen in the course of sample  preparation,\cite{Ralph_private} and  recent 
experiments on hydrogen doped palladium point contacts  indeed exhibit zero bias anomalies associated with 
the presence of hydrogen.\cite{Mihaly} {\gergo However, more detailed calculations would be needed
to estimate the size of the coupling $v_z$ for tunneling hydrogen.}

Tunneling systems with small effective masses can be formed  by dislocations too.\cite{Sethna}
In this case, however, the spatial extent of the defect can be large, and it is not quite clear 
how the two orbital scattering channels driving the two-channel Kondo effect 
could be selected.

Other natural candidates would be impurities with strong magnetic correlations, {\em i.e.} Kondo-like 
impurities. 
In this scenario, two types of Kondo effect take place: a {\em magnetic} Kondo effect with a large Kondo temperature, 
$T_K^{\rm magn}$, and the  {\em orbital} Kondo effect discussed so far at a much smaller energy scale, 
$T_K^{\rm orb }\ll T_K^{\rm magn}$.  
The magnetic  Kondo resonance  would provide the resonance needed for the 
orbital Kondo effect, and the magnetic correlations serve
only to  boost up the couplings of the tunneling system and generate an {\em orbital Kondo effect}
at  lower temperatures.  {\gergo Above the magnetic Kondo temperature, $T>T_K^{\rm magn}$,
magnetic scattering provides also a strong inelastic scattering. Correspondingly, there is a 
temperature-dependent time scale, $\tau_{\rm spin}(T)$  at which the spin of a conduction 
electron is flipped. These spin-flip processes could, in principle,
destroy the two-channel Kondo behavior. To be able to neglect these spin flip 
processes one needs to satisfy the criterion $1/\tau_{\rm flip}(T) < T$ at all temperatures. 
Luckily enough, the rate  $1/\tau_{\rm flip}(T)$ is suppressed below the Kondo scale. 
Unfortunately, however, to our knowledge, this spin flip rate has never been 
determined so far reliably. However, one can obtain a simple estimate for it using the knowledge  
about the inelastic scattering cross section.\cite{FLT} From these considerations one concludes that 
well below $T_K$, in the Fermi liquid range  the spin flip rate must scale as 
$\sim T^2/T_K$, while  in the vicinity of $T_K$ it must be of the order of  
$1/\tau_{\rm spin}(T) \sim T$.  Above $T_K$ this rate must scale to zero logarithmically. 
From these considerations it appears that the criterion $1/\tau_{\rm flip}(T) < T$ is always 
satisfied, and one can therefore probably safely neglect spin flip scattering processes.}
We have to emphasize though that 
these impurities do not need to be Kondo impurities in the usual sense, since the width of the magnetic 
'Kondo resonance' can be in the range of thousands of Kelvins or even larger. For such a 
correlated impurity the 
local density of states remains unrenormalized.\cite{Hewson} However, there is a strong 
{\em field  renormalization} proportional to the $Z$-factor, $Z\sim T^{\rm magn}_K/E_F$, 
which ultimately 
rescales $v_z$ as  $v_z\to v_z /\sqrt{Z}$. We emphasize again that even an extremely large  (magnetic) Kondo 
temperature   in the range of $\sim 1000-10000 {\rm K}$ could give rise to the phenomena discussed 
in this paper,  and therefore the usual  transport and specific heat  anomalies associated with the 
 magnetic Kondo effect may be hardly observed in this case. 

In the case of a tunneling system with strong magnetic correlations, the zero bias anomaly may be very sensitive 
to the external field too. The reason is that in this case the density of states 
and thus the couplings in the two spin channels may depend  rather sensitively on the applied magnetic field. 
This translates to a channel anisotropy in the effective two-channel Kondo model and drives the system to 
a Fermi liquid. This may possibly explain the strong magnetic field dependence in 
some experiments.\cite{RalphReview98}

Strongly coupled electron-phonon systems have also been proposed as possible candidates to produce 
an orbital Kondo effect.\cite{Kusunose} In this case the two-level systems form dynamically.
It is, however, not clear how the regime of extremely strong electron-phonon coupling needed can be reached. 
Furthermore, in these studies only a few vibrational modes could be considered, which has been shown to 
be insufficient to produce the correct low-temperature behavior.\cite{aleiner,borda} 

Finally, let us  comment on the presence of the splitting $\Delta_z$. 
In our previous analysis we completely neglected the asymmetry of the tunneling centers.
In lattice structures such as in Refs.~[\onlinecite{cichorek01,katayama87,brandt82}]
$\Delta_z$ can be rather small. However, in a disordered point 
contact $\Delta_z$ has a random distribution and should provide a low energy 
cutoff for the non-Fermi liquid properties similar to the spontaneous tunneling. 
Therefore a break-down of universal scaling is expected due to the presence of $\Delta_z$. 
Such  a break-down of universal scaling has indeed been observed in Ti point contacts,\cite{Upadhyay97}
where the zero bias anomaly did not depend on the presence of an external magnetic field, 
was sensitive to electro-migration, and had an amplitude 
consistent with the presence of just a few tunneling centers.
These experiments seem to be in  perfect agreement with all  predictions 
of the two-level system model. However, it is much harder to explain the origin of the zero bias anomaly in 
Cu samples.\cite{RalphReview98} While the sensitivity to the magnetic field 
could be explained assuming that the tunneling impurities have a sharp resonance, 
the zero bias anomaly in these experiments has a very large amplitude, 
and as pointed out by Smolyarenko and Wingreen,\cite{smolyarenko} 
the observed anomalous resistivity exponent close to 1/2 can hardly 
be understood assuming a completely random distribution of $\Delta_z$.


I am  grateful to B. Altshuler, I. Aleiner, L. Borda, Sz. Csonka,  A. Halbritter, and A. Zawadowski for 
valuable discussions.   This research has been supported by Hungarian Grants 
Nos. T046267, T046303, and T038162.



\begin{references}
\bibitem{RalphReview98} 
D.C.\ Ralph and R.A.\ Buhrman, Phys.\ Rev.\ Lett.\ {\bf 69}, 2118 (1992);  D.C.\ Ralph, 
A.W.W.\ Ludwig, Jan von Delft, and R.A.\ Buhrman, Phys.\ Rev.\ Lett.\ 
{\bf 72}, 1064 (1994); J. von Delft, 
D.C. Ralph, R.A. Buhrman, A.W.W. Ludwig, and V. Ambegaokar, 
Ann. Phys. (NY) {\bf 263}, 1 (1998). 
\bibitem{Upadhyay97} S.K. Upadhyay, 
R.N. Louie, and R.A. Buhrman,
Phys. Rev. B {\bf 56}, 12033 (1997).
\bibitem{Keijsers96} R.J.P. Keijsers, 
O.I. Shklyarevskii, and H. van Kempen,
Phys. Rev. Lett. {\bf 77}, 3411  (1996).
\bibitem{Balashkin98}
O.P. Balkashin,
R.J.P. Keijsers, H. van Kempen,Yu.A. Kolesnichenko,O.I. Shklyarevskii,
Phys. Rev. B {\bf 58}, 1294 (1998).
\bibitem{Balashkin01}
O.P. Balkashin, I.K. Yansbon, A. Halbritter, G. Mihaly, Fiz. Nizk. 
Temp. {\bf 27}, 1386 (2001) [Sov. J. Low Temp. Phys. {\bf 27}, 1021 (2001)].
\bibitem{cichorek04} See e.g.  T. 
Cichorek, Z. Henkie, J. Custers, {\em et al.},
J. Magnetism and Magnet. Mat. {\bf 272-76}, 66 (2004);
T. Cichorek, H. Aoki, J. Custers, {\em et al.}
Phys. Rev. B {\bf 68}, 144411 (2003).
\bibitem{cichorek01} 
Z. Henkie, A. Pietraszko, A. Wojakowski, L. Kepinski, T. Cichorek, 
J. of Alloys and Compounds {\bf 317-318}, 52 (2001).
\bibitem{cichorek_unpub} 
T. Cichorek, A. Sanchez, P. Gegenwart, {\em et al.}, 
Phys. Rev. Lett. {\bf 94}, 236603 (2005).
\bibitem{katayama87} 
S. Katayama, S. Maekawa, F. Fukuyama, J. Phys. Soc. Jpn. {\bf 56}, 697 (1987). 
\bibitem{brandt82}
N.B. Brandt  S.V. Demishev, V.V. Moshchalkov, S.M. Chudinov,
Fiz. Te. {\bf 15}, 1834 (1981). [Sov. Phys. Semicond. {\bf 15}, 1067 (1982)].
\bibitem{AltshulerComment} 
N. S. Wingreen, B. L. Altshuler, and Y. Meir, 
Phys. Rev. Lett. 75, 769 (1995); N. S. Wingreen, B. L. Altshuler, and Y. Meir, 
Phys. Rev. Lett. 81, 4280(E) (1998);
D. C. Ralph, A. W. W. Ludwig, J. von Delft, and R. A. Buhrman
Phys. Rev. Lett. 75, 770 (1995).
\bibitem{vladzaw} 
A. Zawadowski.  Phys. Rev. Lett. 45, 211-214 (1980);
K. Vlad\'ar and A. Zawadowski, Phys. Rev. B {\bf 28}, 1564 (1983);
{\it ibid} {\bf 28}, 1582 (1983).
\bibitem{Black} For a review see J.L. Black, in: H. G\"untherodt, 
H. Beck (Eds.), Metallic Glasses, Springer, New York, 1981.
\bibitem{Kondo} 
J. Kondo, Physica (Amsterdam) {\bf 84B} 207 (1976). 
\bibitem{Cox} For a review see  
D.L. Cox, A. Zawadowski, Adv. Phys. {\bf 47}, 604 (1998).
\bibitem{Kagan} Yu.M. Kagan and N.V. Prokof'ev, Zh. Eksp. Teor. Fiz. {\bf 90},
2176 (1986)[Sov. Phys. JETP {\bf 63}, 1276 (1986)].
\bibitem{aleiner} I.L. Aleiner,
B.L. Altshuler, Y.M. Galperin, and T.A. Shutenko,
Phys. Rev. Lett. {\bf 86}, 2629 (2001).
\bibitem{aleiner2}
I.L. Aleiner and D. Controzzi, Phys. Rev. B {\bf 66}, 045107 (2002).
\bibitem{EmeryKivelson} V.J. Emery and S. Kivelson, Phys. Rev. B {\bf 46}, 10812 (1992).
\bibitem{VladZimZaw} 
G.T. Zim\'anyi, K. Vlad\'ar, and A.
Zawadowski, Phys. Rev. Lett {\bf 56}, 286 (1986).
\bibitem{ZarZawPRL} G. Zar\'and, Solid State Comm. {\bf 86},
413 (1993); G. Zar\'and, A. Zawadowski, Phys. Rev. Lett. {\bf 72}, 542 (1994).
\bibitem{YuvalAnderson} G. Yuval and P. W. Anderson, Phys. Rev. B {\bf 1}, 1522 (1970).
\bibitem{MoustakasFisher} 
A. Moustakas and D. Fisher, Phys. Rev. B.  {\bf 55}, 6832 (1997).
\bibitem{footnote} In fact, we can prove that within the potential scattering model
of the tunneling center used in Refs.~\protect{\onlinecite{vladzaw,aleiner}} 
where the tunneling impurity interacts with free electrons,
one is never able to get to the other side of the Emery-Kivelson 
line if one assumes as usual only a simple  $s$-wave scattering off the impurity.
Thus the  two-channel Kondo behavior never shows up  in that model. 
\bibitem{borda}  L. Borda, A. Zawadowski, and G. Zar\'and,
 Phys. Rev. B {\bf 68}, 045114 (2003).
\bibitem{Hydrogen} M.J. Puska and R.M. Nieminen, Phys. Rev. B {\bf 27}, 6121 (1983).
\bibitem{Ralph_private} D.C. Ralph, private communication. 
\bibitem{Mihaly} Sz. Csonka {\em et al.},  Phys. Rev. Lett. {\bf 93}, 016802 (2004).
\bibitem{Sethna} T. Vegge {\em et al.}, Phys. Rev. Lett. {\bf 86}, 1546 (2001). 
\bibitem{Hewson} A. Hewson, {\em Kondo model to heavy fermions}, (Cambridge University Press, 1993). 
\bibitem{smolyarenko}  I. E. Smolyarenko and N. S. Wingreen
 Phys. Rev. B {\bf 60}, 9675-9689 (1999).
\bibitem{Kusunose} H. Kusunose and K. Miyake, J. Phys. Soc . Jpn. {\bf 65}, 3032 (1996); 
M. Kojima, S. Yotsuhashi, K. Miyake,  Acta Phys. Pol. B {\bf 22}, 1331 (2003).
\bibitem{Nozieres} 
A. Zawadowski, G. Zar\'and, P. Nozi\`eres, K. Vlad\'ar, and G.T. Zim\'anyi,
Phys. Rev. B {\bf 56}, 12947 (1997).
\bibitem{FLT} 
Nozi\`eres P., J. Low Temp. Phys. {\bf 17}, 232 (1974); for a recent accurate calculation
see G. Zar\'and, L. Borda, J. von Delft, and N. Andrei, Phys. Rev. Lett. {\bf 93}, 107204 (2004).
\bibitem{Zarunpub}
G. Zar\'and, unpublished.
\end{references}
\end{document}